\documentclass[fleqn,usenatbib]{mnras}

\usepackage{newtxtext,newtxmath}

\usepackage[T1]{fontenc}
\usepackage{ae,aecompl}

\usepackage{graphicx}    
\usepackage{amsmath}    
\usepackage{amssymb}    

\title[Test of the pc-GR theory with EHT observations]
{Predictions of the pseudo-complex theory of Gravity for
EHT observations: \\
I. Observational tests}

\author[Th. Boller et al.]{
Th. Boller,$^{1,4}$\thanks{E-mail: bol@mpe.mpg.de}
P.O. Hess,$^{2,4}$
A. M\"uller$^{3}$,
and H. St\"ocker$^{4,5,6}$
\\
$^{1}$Max-Planck-Institute for extraterrestrial physics, Giessenbachstrasse 1, 85741 Garching, Germany\\
$^{2}$Instituto de Ciencias Nucleares UNAM, C.U., Circuito exterior S/N
A.P. 70-543, 04510 Mexico D.F., Mexico\\
$^{3}$ Excellence Cluster Universe, Technical University Munich, Giessenbachstrasse 1, 85741 Garching, Germany\\
$^{4}$Frankfurt Institute for Advanced Studies, Johann Wolfgang Goethe
 Universit\"at, Ruth-Moufang-Str.1, 60438 Frankfurt am Main, Germany \\  
$^{5}$Goethe Universit\"at,
Max-von-Laue-Str. 1, 60438 Frankfurt am Main, Germany \\
$^{6}$GSI Helmholtzzentrum f\"ur Schwerionenforschung GmbH, Planckstrasse 1, 64291 Darmstadt, Germany
}

\date{Accepted 2019 February 21. Received 2019 February 04; in original form 2018 December 26.}

\pubyear{2019}

\begin{document}
\label{firstpage}
\pagerange{\pageref{firstpage}--\pageref{lastpage}}
\maketitle

\begin{abstract}
A modified theory of gravity, avoiding singularities in the standard theory of gravitation, has been developed by Hess \& Greiner, known as the pseudo-complex theory of gravitation. The pc-GR theory shows remarkable observational differences with respect to standard GR. The intensity profiles are significantly different between both theories, which is a rare phenomenon in astrophysics. This will allow robust tests of both theories using Event Horizon Telescope (EHT) observations of the Galactic Center. We also predict the time evolution of orbiting matter. In this paper we summarize the observational tests we have developed to date. The theory is described in the second paper of this series (Hess et al. 2019, referred to as paper II hereafter).
\end{abstract}


\section{Introduction}

Einstein's General Relativity (GR), which was developed 100 years ago, successfully describes gravitation and has withstood all four classical experimental tests, (i) the precession  of Mercury, (ii) the Shaprio delay, (iii), the deflection of light, and (iv) gravitational  wave emission from   Hulse-Taylor pulsars. Beyond these a set of other measurements to test Einstein's  theory have been critically assessed by 
\citet{2016AAS...227.9001W} in the strong gravity regime and in gravitational waves.

Motivated by the existence of singularities in Einstein's General Theory of Gravity, e.g. the curvature singularity at $\rm r=0~m$
\citet{1916SPAW.......189S}
(Kretschman-scalar), the orbital velocity of v=c at the Schwarzschild radius of infalling matter into the central black hole,
and the fact that observations from external observers inside the Schwarzschild Horizon are not possible, excluding regions in the Universe from observation, 
\citet{2009IJMPE..18...51H} 
published a new algebraic extension of the standard GR, called pseudo-complex General Relativity (pc-GR).

One of the important consequences of the pc-GR theory is the presence of a field with repulsive properties.
This prevents gravitational collapse for very large masses
\citep{2010IJMPD..19.1217H, 2015_book_Hess}.
PC-GR predicts that the
emission from orbiting
matter has  different  timing  characteristics than that which would be predicted by standard GR  due  to  the  different  values  of the  gravitational redshift and orbital frequency 
\citep{2013eipq.book..293B, 2014eipq.book..254B, 2014MNRAS.442..121S}.

The fall-off dependence of the dark-energy density as a function of $r$ is given by the
parameter n (c.f. Eq. 1 in paper II). 
For a value of n = 3 the dark-energy density versus radius dependency does not contradict the tests of standard GR in the solar system.
\citet{2018AN....339..298N} have pointed out that a larger value of n is required to account for the results obtained from gravitational wave observations (c.f. their Eq.4).
Paper II shows that for larger values of n, the predicted
intensity and velocity structures will be almost 
identical within pc-GR,
with only slight changes in the absolute values. 
For instance, increasing n from 3 to 4 results in a shift in the maximum orbital frequency from 1.72 m to 1.89 m and a shift in the maximum orbital frequency from 0.21 to 0.22 $\rm \frac{c}{m}$ (c.f. Eq. 4 of paper II and  Fig.~\ref{fig:simu_2} in this paper). Therefore the appearance of a dark and a bright ring in the intensity profiles discussed in Section~\ref{obs_tests} are robust and will not change for steeper values of n. 
In this paper we assume a value of n=4 
for the tests on the timing properties to account for the detection of gravitational waves. We assume a value of n=3 for the ray-tracing simulations, as the structures in the intensity behaviour will not change.

The Galactic Center has been studied in great detail with near-infrared observations.
Based on the orbital motions of stars in the vicinity of the Galactic Center, the mass of the central black hole has been constrained to $\rm (4.31\pm0.06)\times 10^6\ M_{\sun}$
(\citet{2009ApJ...692.1075G}, see also
\citet{1996ApJ...472..153G, 1997MNRAS.284..576E}).
%
%
%
\citet{2018A&A...615L..15G} 
have, for the first time, measured the 
gravitational redshift of the star S2 near the Galactic Center.
In addition, 
\citet{2018A&A...618L..10G}
have detected
bright flares at distances of approximately  {\rm $150~ \mu$as} from the central black hole.

In this paper we summarize our imaging and timing predictions based on the pc-GR theory for the Galactic Center.
In Section~\ref{imaging} the ray-tracing simulations are summarized.
In Section~\ref{timing} we predict the orbital velocities of particles which can be compared to EHT observations at different epochs.
Finally in Section~\ref{profiles} we compare the emission profiles predicted by pc-GR and standard GR, which show remarkable differences. The theory of pc-GR in general, along with a specific focus on the tests discussed in this work, is presented in 
\citet{2019...HessBoller} 
(hereafter Paper II: Theory and robust predictions).

\section{Observational tests}\label{obs_tests}

In this Section we describe the observational tests we have developed to date to verify the predictions of the pc-GR theory in preparation for the upcoming EHT observations.
These tests include ray-tracing simulations
of the flow of accreting matter around black holes
using both standard GR and pc-GR.
In addition, we have produced diagnostic plots comparing both theories with regard to emissivity profiles and the timing properties of orbiting matter.
The main results are described in
\citet{2014eipq.book..254B,2014MNRAS.442..121S,2016grph.book..111S,2018eipq.book..999B}.

\begin{figure}
    \includegraphics[width=\columnwidth]{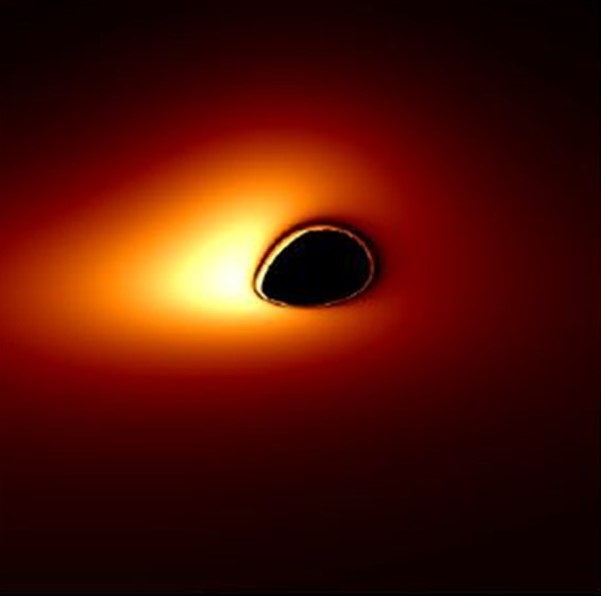}
    \caption{PC-GR ray tracing simulation of the emission around a rotating black hole in the Galactic Center. The new features are the appearance of a black ring at 1.72 m and a bright ring at smaller distances.  
    The simulations are done for n = 3 as the structures will not change for steeper slopes of the energy-density dependence on distance.
        }        
    \label{fig:simu_1}
\end{figure}


\begin{figure}
    \includegraphics[width=6.1cm,angle=-90]{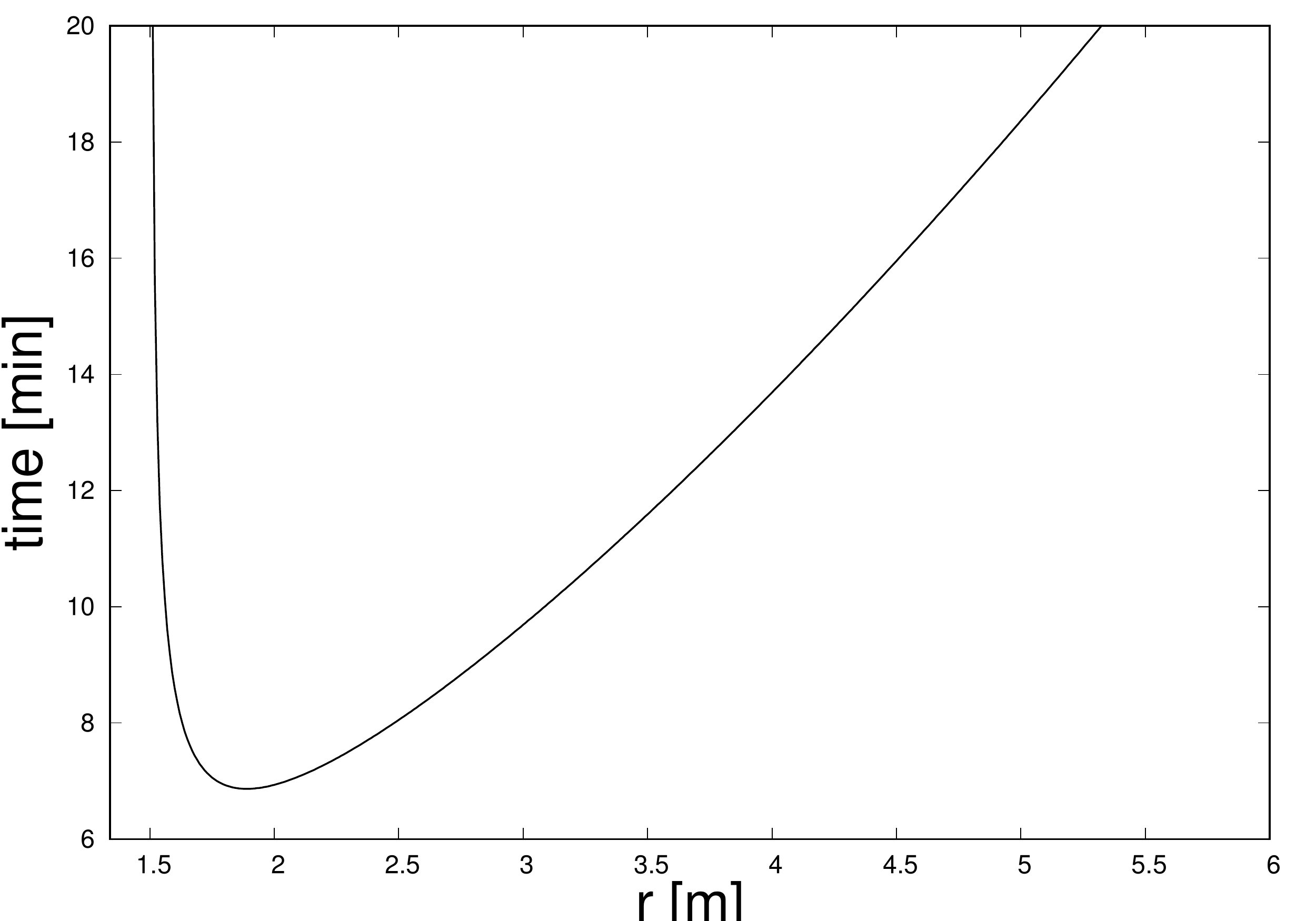}
    \caption{Orbital time in units of minutes as a function of distance.
    Note that in contrast to the standard GR theory there exists a minimum orbital time scale.
    }
    \label{fig:simu_2}
\end{figure}


\subsection{Ray-tracing simulations and observational predictions}\label{imaging}

Intensive ray-tracing simulations of accretion onto the Galactic Center have been performed by 
\citet{2014MNRAS.442..121S} 
for both the standard GR and pc-GR theories. Since the gravitational redshift predicted by pc-GR is less than that predicted by standard GR, the emission from pc-GR accretion is typically brighter.

Specifically, the intensity profiles of the emitting material predicted by both theories is significantly different for spin values above about 0.5 (see Figs. 3 and 4 of \citet{2014MNRAS.442..121S}),
which is the case for the Galactic Center 
\citep{2003Natur.425..934G}.

In Fig.~\ref{fig:simu_1} we show a pc-GR simulation of the emission around the rotating (a = 0.6 m) black hole in the Galactic Center.
For distances greater than 1.72 m (1.89 for n=4), the standard blue-shifted emission from infalling matter is similar to what is predicted for a standard GR black hole, but with a higher intensity.

The orbital time scale exhibits a minimum of 6.81 minutes (see Fig.~\ref{fig:simu_2}).
The first derivative of orbital frequency is proportional to the flux. Therefore, at the minimum orbital time scale the flux is equal to zero. This we refer to as the dark ring. This is a new feature not predicted by standard GR which may be resolved by EHT observations.

At distances smaller than 1.72 m, the orbital frequency  decreases as the flux increases;
this leads to what we refer to as the inner bright ring.


\begin{figure}
    \includegraphics[width=6.0cm,angle=-90]{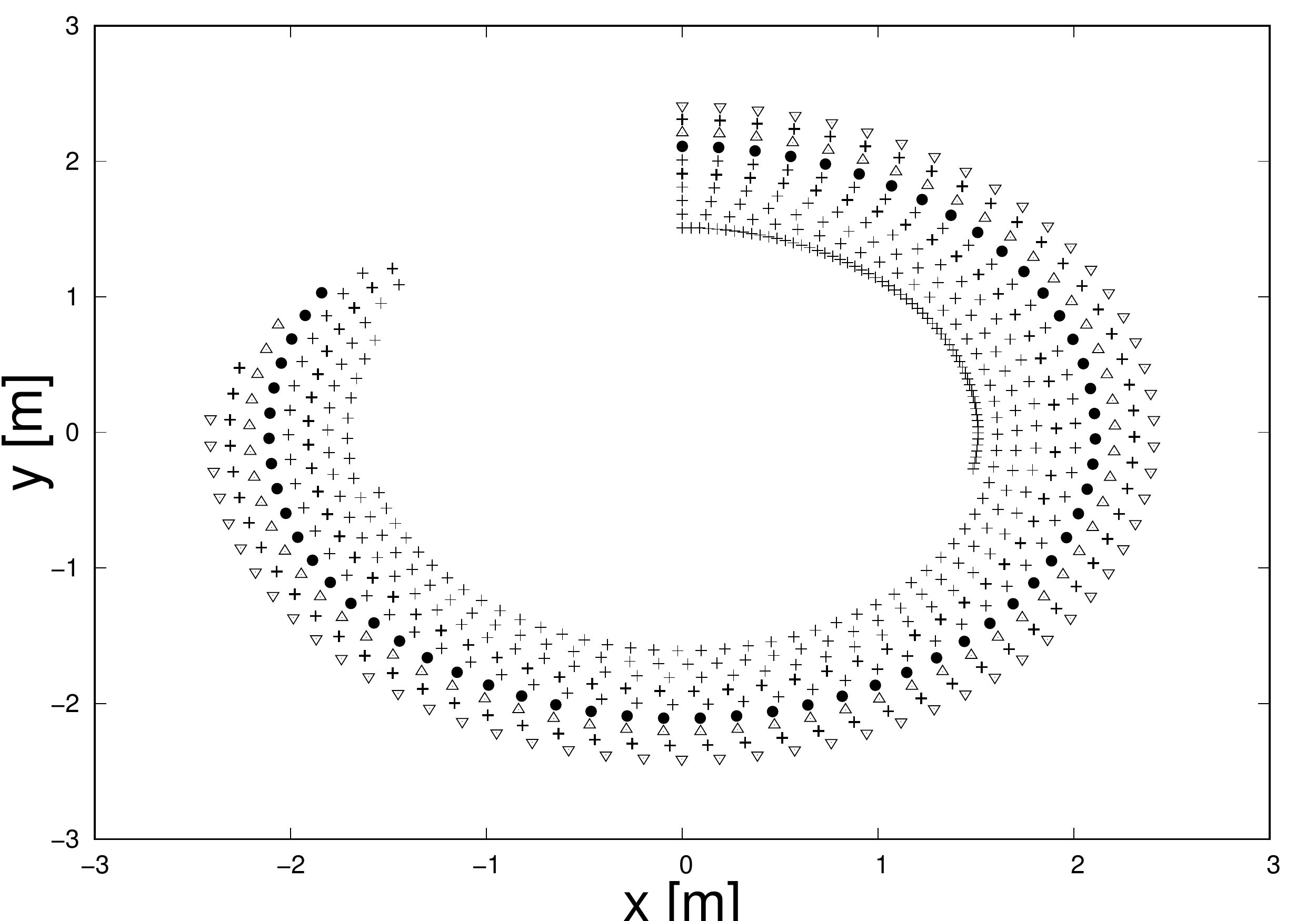}
    \includegraphics[width=6.0cm,angle=-90]{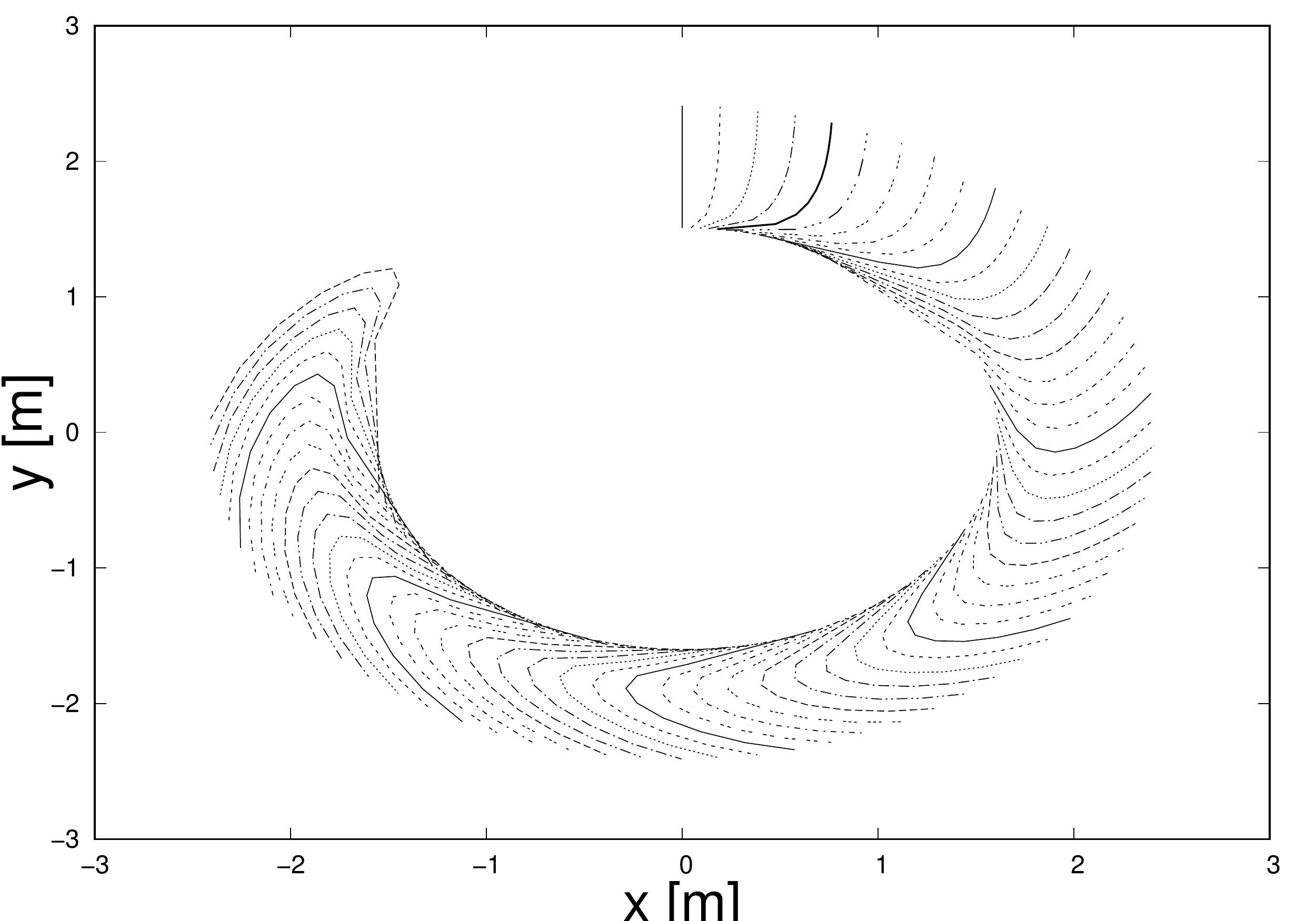}
    \caption{Orbital motions of particles predicted by the pc-GR theory
    (see Section \ref{timing} for further details).
    The upper panel shows the individual particles at a given distance from the black hole. In the lower panel equal time steps have been connected with a solid line for illustration purposes. With these plots we can predict the motion of particles around the Galactic Center with the upcoming EHT observations. }
    \label{fig:Figure_orbitalmotion1}
\end{figure}

\subsection{Time dependencies, predictions, and tests}\label{timing}

An additional feature of the pc-GR theory which can be tested via EHT observations is the orbital motion of particles moving around the central black hole.
In standard GR the particles reach the
$\rm limit\  v\rightarrow c$
at the Schwarzschild radius, which is a singularity.
This is not the case in pc-GR, as described in Section \ref{imaging} (see also Fig. \ref{fig:simu_2}).
The orbital frequency is given by Eq. 4 in paper II. For n = 4 the maximum orbital frequency 
$\omega_{max}$ is $0.22\ \frac{c}{m}$
corresponding to a minimum orbital time scale
$t_{min}$ of 6.81 minutes.



As the orbital frequency is a known function of distance from the black hole we can predict the orbital motions around the  black hole. This is illustrated in
Fig.~\ref{fig:Figure_orbitalmotion1}.
The upper panel shows the movement of particles, assuming for illustration purposes a radial infall at $\phi$ =0 degrees and an infinitesimal width. According to  Fig.~\ref{fig:simu_2} particles at 
1.89 m
move with the highest orbital frequency.
The time steps are 6 seconds for a total length of 6 minutes.
The closest orbit is at 
1.5 m (c.f. Eq. 3 in paper II) 
and the orbits are separated by 0.1 m.
In total we show 10 curves, and the largest distance is at 
2.24~m.
At the highest orbital frequencies a full orbit takes 
6.21 minutes.
The innermost orbit  moves with the lowest velocity, and the fastest orbit is at 
1.89~m
which has not been closed.

 In general, any spherical infall of matter onto the central black hole will become smeared out, and a V-shaped emission pattern will appear with a peak in the emission at the fastest orbit and an inner distribution which is spread out more than the outer arm.

In the lower panel of Fig.~\ref{fig:Figure_orbitalmotion1}
we connect particles moving at the same time steps.
The solid lines in the inner part of the Figure arise as some particles have moved such large distances that they have become connected.

\subsection{Emission profile predictions}\label{profiles}

A  robust test of the pc-GR theory can be made by examining the flux distribution as a function distance from the central black hole.

\begin{figure}
    \includegraphics[width=\columnwidth]{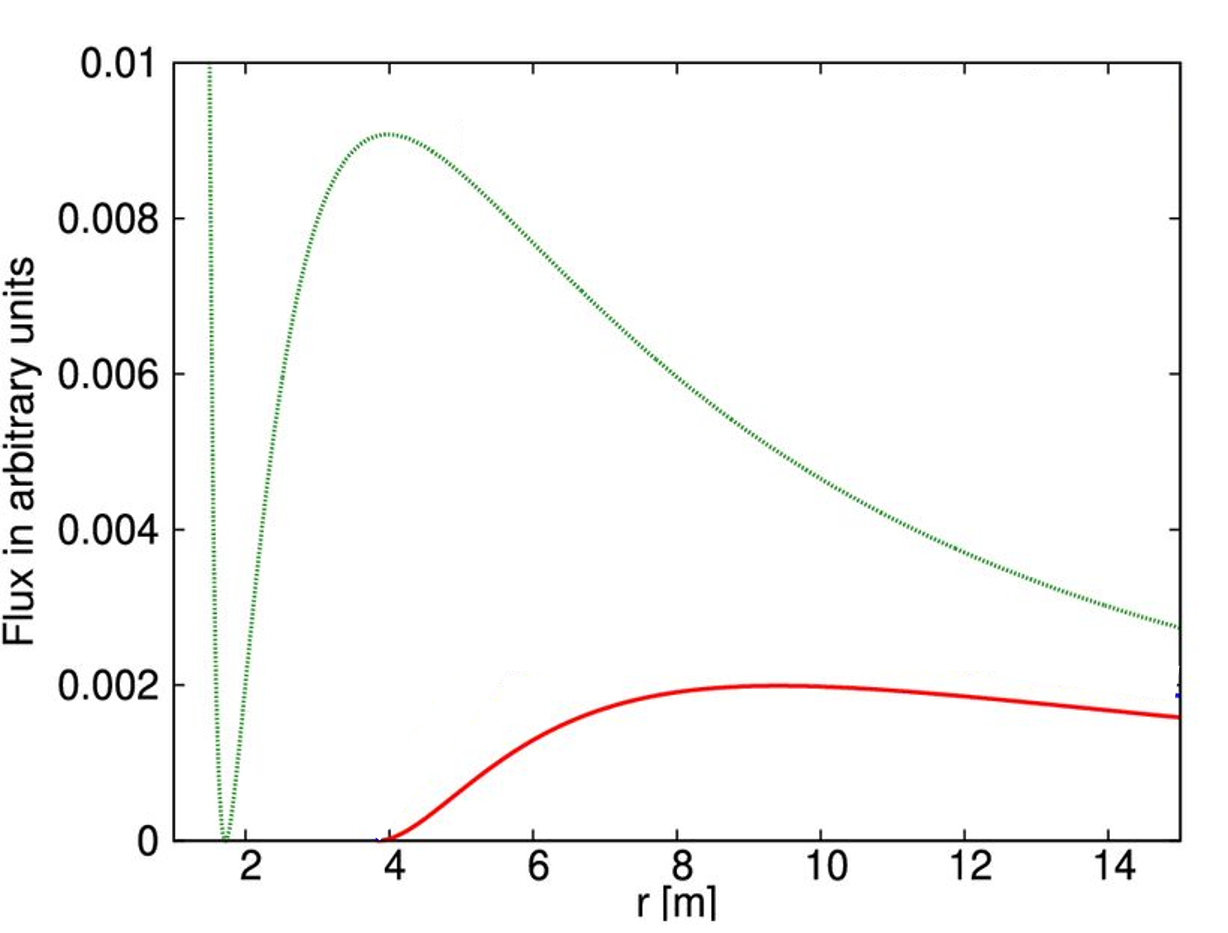}
    \caption{Flux emitted by infalling matter onto the black hole as a function of distance for the standard GR (lower red line) and the pc-GR theory (upper green line
    and n=3). 
    The significant differences in flux will allow a robust test of the pc-GR prediction. The curves are calculated for the mass of the black hole in the Galactic Center and a spin of a = 0.6 m.}    
    \label{fig:Figure_Kurvendiskussion}
\end{figure}

Fig.~\ref{fig:Figure_Kurvendiskussion} 
(adapted from Fig. 2 of 
\citet{2014MNRAS.442..121S}) 
shows the flux in arbitrary units for the standard GR theory (in red) and the pc-GR theory (in green) as a function of distance from the black hole (calculated for the mass of the black hole in the Galactic Center and a spin of a = 0.6 m).

For the standard GR theory the flux increases up to about
8 m, and decreases towards smaller distances from the black hole due to the increasing gravitational redshift, vanishing at about 4 m.
In contrast, the pc-GR theory predicts an increasing in flux up to about 4 m, due to the reduced gravitational redshift and the repulsive forces.
The flux decreases towards 1.72 m (1.89 m for n = 4), reaching  zero intensity at these values. Below 1.72 m the orbital frequency sharply decreases, resulting in emission refereed to as the inner bright ring.

This flux distribution is remarkably different to standard GR theory, which is a rare phenomenon in astrophysics. 
The differences between n=3 and n=4 are small in absolute values and the intensity structures do not change.
If the EHT observations will resolve the emission below about 10 m,
then it may be possible conduct crucial tests of pc-GR theory.

\section{Conclusions}

We have described how
the pc-GR theory of gravitation can be tested against the standard GR theory via the upcoming EHT observations. 
The two theories predict remarkably different emission profiles for infalling material around a black hole.
We also predict the time scales of orbiting matter, which can be compared with time dependent EHT images. If our imaging and timing predictions will be confirmed by EHT observations, this will give us further motivation to investigate our Ansatz of a modified GR theory without singularities, giving late tribute to the prophetic view of Prof. Walter Greiner.

\section*{Acknowledgements}

TB and PH are grateful for the longstanding support and collaboration with the Frankfurt Institute of Advanced Studies. The authors thank Damien Coffey for a critical reading of the manuscript. POH thanks for the financial help from DGAPA-PAPIIT (IN100418).
H.St. acknowledges the Judah M. Eisenberg Professor Laureatus at the Fachbereich Physik endowed by the Walter Greiner-Gesellschaft zur F\"orderung der physikalischen Grundlagenforschung e.V. Frankfurt am Main.
The authors would like to thank the referee whose comments greatly improved the content of this paper.



\bibliographystyle{mnras}
\bibliography{mnras_boller-I-r2-final}

\bsp    
\label{lastpage}
\end{document}